\documentclass[aps,prl,showpacs,preprint,superscriptaddress]{revtex4}
\usepackage{graphicx,color}

\begin{document}

\title{Ultrarelativistic nanoplasmonics as a new route towards extreme intensity attosecond pulses}

\author{A.A.~Gonoskov}
\affiliation{Institute of Applied Physics, Russian Academy of Sciences,
603950 Nizhny Novgorod, Russia}
\affiliation{Department of Physics, Ume{\aa} University, SE-901 87 Ume{\aa},
Sweden}

\author{A.V.~Korzhimanov}
\affiliation{Institute of Applied Physics, Russian Academy of Sciences,
603950 Nizhny Novgorod, Russia}
\affiliation{Department of Physics, Ume{\aa} University, SE-901 87 Ume{\aa},
Sweden}

\author{A.V.~Kim}
\affiliation{Institute of Applied Physics, Russian Academy of Sciences,
603950 Nizhny Novgorod, Russia}

\author{M.~Marklund}
\affiliation{Department of Physics, Ume{\aa} University, SE-901 87 Ume{\aa},
Sweden}

\author{A.M.~Sergeev}
\affiliation{Institute of Applied Physics, Russian Academy of Sciences,
603950 Nizhny Novgorod, Russia}

\date{\today}

\begin{abstract}
The generation of ultra-strong attosecond pulses through laser-plasma interactions offers the opportunity to surpass the intensity of any known laboratory radiation source, giving rise to new experimental possibilities, such as quantum electrodynamical tests and matter probing at extremely short scales. Here we demonstrate that a laser irradiated plasma surface can act as an efficient converter from the femto- to the attosecond range, giving a dramatic rise in pulse intensity. Although seemingly similar schemes have been presented in the literature, the present setup deviates significantly from previous attempts. We present a new model describing the nonlinear process of relativistic laser-plasma interaction. This model, which is applicable to a multitude of phenomena, is shown to be in excellent agreement with particle-in-cell simulations. We provide, through our model, the necessary details for an experiment to be performed. The possibility to reach intensities above $10^{26}$~W/cm$^2$, using upcoming 10 petawatt laser sources, is demonstrated. 
\end{abstract}


\maketitle
\paragraph{Introduction. ---} Recent progress in ultrahigh-power laser technology has resulted in pulse intensities surpassing 10$^{22}$ W/cm$^2$ \cite{yanovsky.oe.2008}, and stimulated the construction of multi-petawatt laser sources \cite{ELI}. Such lasers open up opportunities for studying both a number of fundamentally new problems, such as the effects of vacuum nonlinearities \cite{bell.prl.2008, dunne.prd.2009, fedotov.prl.2010} in laser fields and photo-nuclear physics, as well as some very important applications, e.g., laser based particle acceleration, fast ignition fusion schemes, and the generation of electromagnetic radiation with tailored properties.

Given this, the study of overdense plasmas irradiated by relativistically intense laser pulses constitutes a very important and challenging research direction. Numerical studies using particle-in-cell approach, taking into account most important effects for the typical range of parameters, are known to be an excellent tool in this field, and the numerical results in general agree well with the experimental results. Furthermore, so called nonlinear fluid models \cite{akhiezer.jetp.1956} give a set of equations analytically describing such processes, but the strongly nonlinear plasma behavior, due to the ultrarelativistic motion of the plasma electrons, makes the development of a theoretical approaches a highly complex task. Thus, one is normally forced to limit oneself to qualitative analyses, and use a phenomenologically motivated ad hoc treatment.

The generation of high harmonics from intense laser-plasma interactions is an intensely studied research field, with a manifold of applications \cite{teubner.rmp.2009}, including the idea of reaching the extreme intensities needed to probe vacuum nonlinearities using lasers \cite{bulanov.prl.2003, naumova.prl.2004, gordienko.prl.2005, quere.prl.2006, tarasevich.prl.2007}.

As of today, the most prominent theoretical model used in the analysis of such high-order harmonics generation (HHG) is the so-called \emph{oscillating mirror model} (OMM). In the OMM, one considers the backradiation from an overdense plasma by taking into account the retarded emission from the oscillating source. This approach was first proposed by Bulanov et al. \cite{bulanov.pp.1994} and developed further in Refs. \cite{lichters.pp.1996, dvonderlinde.apb.1996}. Lately, the OMM approach has been reexamined by Gordienko et al. \cite{gordienko.prl.2004}, who proposed that at each moment of time there exists a so-called \emph{apparent reflection point} (ARP) at which the energy flux vanishes.  This assumption implies a local (in time) energy conservation or, phrased differently, the approach neglects the energy accumulated by the plasma, in the form of the fields due to charge separation caused by the light pressure. An asymptotic analysis of the ARP dynamics in the strongly relativistic limit \cite{baeva.pre.2006} indicates the universal properties of the HHG spectra: the intensity of $n$th harmonic scales as $n^{-8/3}$, and the cutoff $\sim \gamma_{\rm max}^3$, where $\gamma_{\rm max}$ is the maximal relativistic factor of the ARP. These results agree well with the experiments \cite{dromey.nphys.2006, nomura.nphys.2009, dromey.nphys.2009}.

\begin{figure}
\includegraphics[width=0.9\columnwidth]{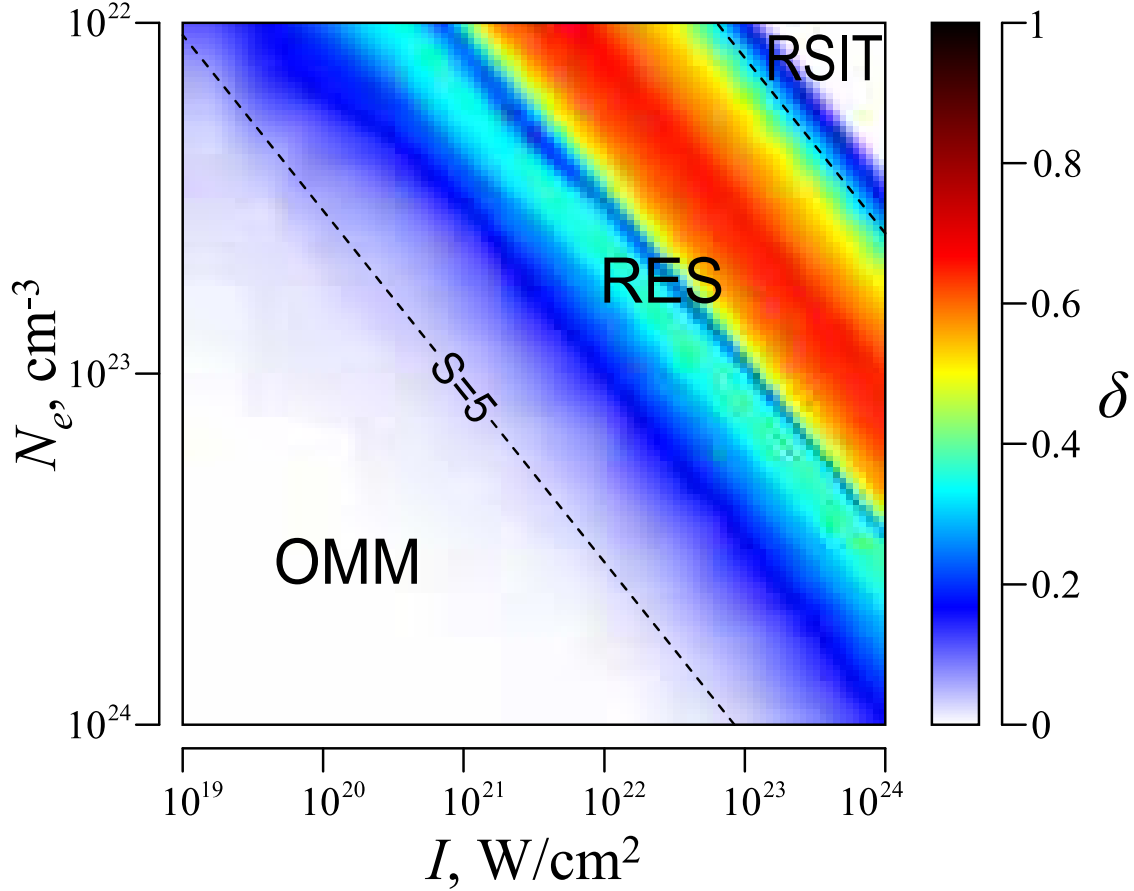}
\caption{The parameter $\delta$ obtained from 1D PIC simulations of a plasma with density $N_e$ at oblique irradiation ($\theta = 60^\circ$, P-polarization) by a wave with constant intensity $I$ and a $1 \mu\mathrm{m}$ wavelength during one optical cycle. ($S$ is the ultrarelativistic similarity parameter, defined as the quotient between dimensionless density and the dimensionless amplitude (see Eq. (2)).)} \label{fig1}
\end{figure}

Nevertheless, the assumption of the OMM concerning local temporal energy conservation is valid only for restricted values of the plasma density and laser intensity.
The parameter $\delta$, used in Fig.~\ref{fig1}, is defined as the ratio between the maximal accumulated plasma field energy and the energy of one optical cycle. In the bottom left corner of the figure we find the zone labeled "OMM", for which $\delta \ll 1$. Thus, here the energy accumulation can be neglected, as assumed in the OMM. In the top right corner the region of relativistically self-induced transparency (RSIT) is shown. Thus, there is a large, and very important, parameter region that so far has not been covered by any theoretical model, and for which $\delta \sim 1$ such that the OMM's assumption of local energy conservation is no longer valid. Here collective ultrarelativistic electrons motion can give rise to a \emph{nanoplasmonic structures}, i.e. nanometer scale surface layers and their oscillations, absorbing at each period the energy of the laser pulse to the energy of internal electric and magnetic plasma fields and then reemmiting this energy in the form of attosecond burst. The process result in a markedly slower decay in the generated higher harmonic spectra \cite{boyd.prl.2008} as compared to the OMM results. Furthermore, based on the phenomenological assumption of electron nanobunches appearing in the plasma, emitting radiation, it was shown in Ref. \cite{anderbrugge.pp.2010}  that the spectra can be much flatter than predicted by the OMM. Moreover, the OMM assumes as a prerequisit that the incident and backradiated amplitudes at the ARP are equal, in accordance to the Leontovich boundary conditions (which is in direct correspondence to the local energy conservation). Consequently, situations where large field amplifications is to be expected cannot be analyzed through this model. Thus, finding a new theoretical model in the relevant parameter regime, between the OMM and RSIT regions, is of the utmost importance for a large number of applications. 

In the present work we propose, for the first time, a physically motivated model, a so-called \emph{relativistic electronic spring} (RES) model, describing the highly nonlinear behavior of laser-plasma interactions. The model gives very good agreement with simulation and makes it possible to analytically study a vast range of regimes in laser-plasma interactions that otherwise would be out of reach for analysis. In particular, one of the most remarkable effects in the RES regime is the possibility of generating attosecond pulses with an amplitude several orders of magnitude higher than the incident laser pulse. Here, we apply the RES model to this amplification effect in order to understand the underlying physical mechanisms and determine the optimal parameters for an experiment to be performed. We compare the results with particle-in-cell simulations and find excellent agreement. The implications of our results are discussed, in particular the possibility to utilize this new type of secondary source for novel experiments.

\begin{figure}
\includegraphics[width=0.9\columnwidth]{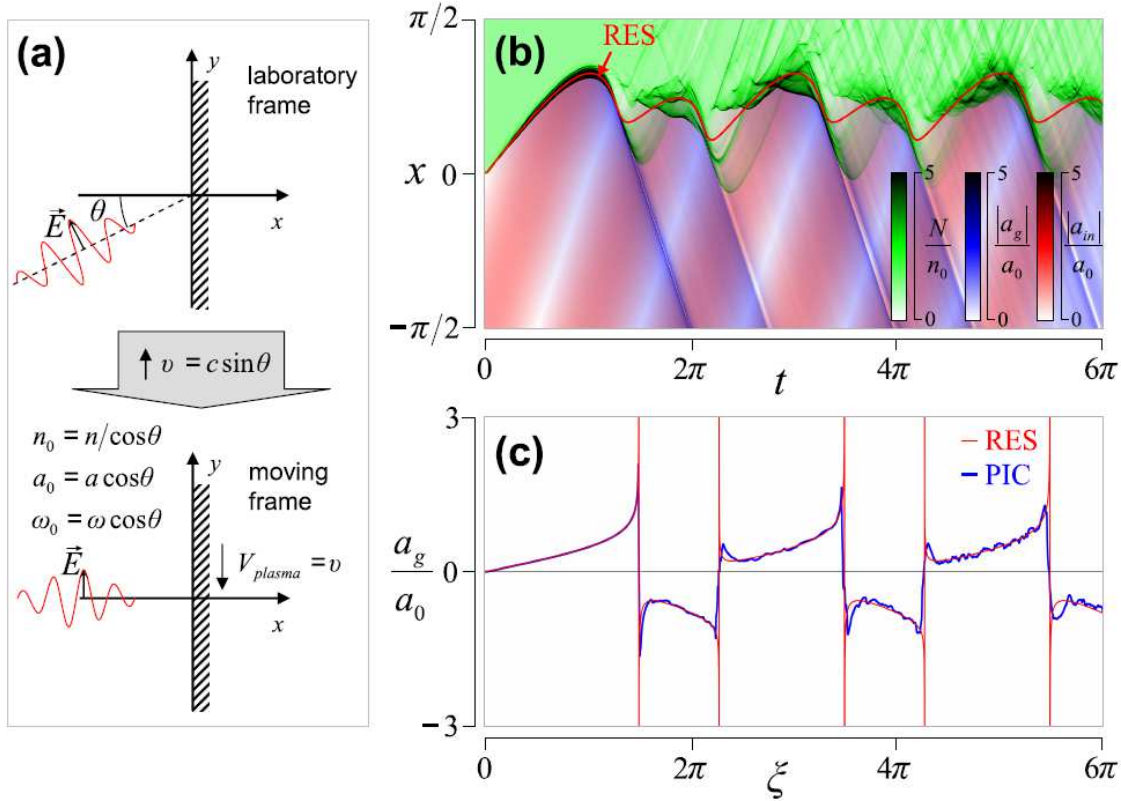}
\caption{(a) Transformation to a moving reference frame. Here, the plasma
density $n$, wave amplitude $a$, and frequency $\omega$ have a subscript $0$
denoting the moving frame. (b) Space-time distribution of electrons density $N$
(green), amplitudes of incident $a_{in}$ (red) and backradiated $a_g$ (blue)
electromagnetic fluxes obtained from 1D PIC simulation of plasma with density $4
\times 10^{23}$ cm$^{-3}$ at oblique irradiation ($\theta = 11.25^{\circ}$) by a
wave with constant intensity $10^{23}$ W/cm$^2$ and 1 $\mu$m wavelength during
three optical cycles. The red curve shows position of the thin layer obtained
using the RES model. Time and coordinate are in dimensionless units (see. Eq.
(\ref{dim})). (c) Backradiated signal obtained by PIC simulation (blue) and
using the RES model (red).} \label{fig2}
\end{figure}

\paragraph{Ultrarelativistic energy conversion on the surface of a planar target. ---} The process of giant attosecond pulse
generation due to the oblique incidence of a p-polarized electromagnetic wave on a
plane plasma boundary $(x=0, y, z)$ may be considered using a boosted frame
moving along the plasma surface and plane of incidence in the $y$-direction with the 
velocity $c \sin \theta$, where $c$ is the speed of light and $\theta$ is the
angle of incidence (see Fig. \ref{fig2}(a)), thus making the problem one-dimensional \cite{Bourdier}.

The incident laser pulse pushes the electrons into the plasma due to the light
pressure. Unlike the case of normal incidence, the oblique incidence results in
the emergence of uncompensated currents and magnetic fields in the boosted
frame. Therefore, the electrons experience an additional ponderomotive action,
which is different during the two half-periods of the incident wave. When the
laser electric field is directed along the $y$-axis, the Lorentz force due to
uncompensated currents enhances the light pressure effect, pushing electrons
further from the boundary. 
It can be analytically shown that under the laser radiation pressure the electrons are shifted and group into a boundary layer. The thickness of this layer tends to zero as the laser intensity increases and, in the ultrarelativistic limit, is much smaller than all other spatial scales involved in the process. The charge and current densities in this layer greatly exceed the ones in the unperturbed plasma. At this stage the incident wave energy is
transformed to the energy of the internal plasma fields and kinetic energy of
the particles.

The formation of an ultrathin (nanoscale) electron layer due to the interaction
between an ultrarelativistic laser pulse and an overdense plasma has been known
for about a decade, being reported in works on relativistic self-induced
transparency \cite{Kim, eremin.pp.2010} and particle acceleration using thin foils
\cite{Gonoskov}. Unlike in the case of a circularly
polarized laser pulse, in which electrons may be shifted by a distance of several
wavelengths, a linearly p-polarized pulse results in a light pressure force that oscillates within a
field period. Hence, the electrons are pushed from the surface for no longer
than a fraction of the optical wavelength, then break away under the action of the 
charge separation force and travel towards the incident wave in the form of a
nanobunch, providing a source of attosecond burst. The presently described processes can clearly be seen in Fig.~\ref{fig2}(b), where the results of 1D PIC simulation
are presented.

Thus, the described process and concomitant energy conversion may be represented
as a sequence of three stages: 1) the pushing of electrons from the surface by the
ponderomotive force and the formation of a thin current layer giving an energy
transfer from the laser field to the plasma fields and particles; 2) the backward
accelerated motion of the electrons towards the incident wave with the conversion of
the energy accumulated in the plasma and laser field energy into kinetic energy
of an ultrarelativistic electron bunch; and 3) the radiation of attosecond
pulses by an electron bunch due to conversion of the kinetic energy and laser
field energy to the XUV and X-ray range. Based on the motion of the plasma
electrons and the energy conversion scenario we find that it is natural to refer
to this three-step process as to a model of a \textit{relativistic electronic
spring} (RES). It should be emphasized that due to the energy accumulation in
the plasma, the backradiated field can be much larger than the incident field. This is the fundamental difference from
the OMM (see, e.g., \cite{mourou.rmp.2006, marklund.rmp.2006}), a model which assumes a direct correspondence between the incident and backradiated field.


\paragraph{The RES model. ---} A model
describing the dynamics of a thin current layer and the generation of attosecond pulses
may be formulated starting from three intuitively clear and 
physically justified prerequisities, that can be verified using PIC simulation. First, we will assume
that at each moment of time the plasma electrons are represented by two
fractions: one infinitely narrow layer of shifted electrons at a certain moving
point $x_s$, where all the electrons from the region $0 < x < x_s$ are
accumulated, and one population of electrons with unperturbed density at $x > x_s$. Second, we
will characterize the motion of particles in the layer by the ensemble averaged
$x$- and $y$-projections of the velocities $\beta_x$ and $\beta_y$ (normalized by
the speed of light, and with the $\gamma$-factor defined by $\beta^2_x + \beta^2_y = 1 - \gamma^{-2}$). Third, we will suppose
that the motion of the electrons in the thin layer together with the flow of
uncompensated ions in the $0 < x < x_s$ region fully compensate the incident
electromagnetic radiation in the unperturbed plasma at $x > x_s$. 

It is readily shown from Maxwell's equations that in a one-dimensional geometry, 
a moving charged layer, with surface charge $\sigma$, emits electromagnetic waves
with amplitudes $2 \pi \sigma \beta_y/\left(1 - \beta_x\right)$ and $2 \pi
\sigma \beta_y/\left(1 + \beta_x\right)$ in the positive and negative directions
of the $x$-axis, respectively. 
Consequently, the expression for
the incident wave compensation motion may be written 
\begin{equation}
	\sin \left( x_s - t \right) = \frac{S}{2\cos^3\theta}\left( \sin\theta -
\frac{\beta_y}{1 - \beta_x}\right)x_s,
	\label{skining}
\end{equation}
where the left-hand side corresponds to the incident wave, whereas the terms on the right-hand side describe the radiation of the uncompensated ions and the electron layer, respectively; here $S = n/a$ is the ultrarelativistic similarity parameter, while $a$ and $n$ are the incident wave amplitude and the plasma density in the laboratory frame, respectively. We use dimensionless quantities which can be expressed in terms of dimensional time $\hat{t}$, coordinate $\hat{x}$, density $\hat{n}$ and electric field amplitude $\hat{a}$ according to
\begin{equation}
	t = \omega\cos\theta\hat{t},~x = \frac{\omega\cos\theta}{c}\hat{x},~n =
\frac{4 \pi e^2}{m \omega^2}\hat{n},~a = \frac{e}{m c \omega}\hat{a},
	\label{dim}
\end{equation}
where $\omega$ is the carrier laser frequency, while $m$ and $e$ are the electron mass
and charge, respectively. 
Analogous to the Eq. (\ref{skining}), the electric field, as a function of retarded time $\xi = x + t$, emitted by the plasma in the negative $x$-direction is given by 
\begin{equation}
	a_g \left[\xi = x_s(t) + t \right] =
a_0\frac{S}{2\cos^3\theta}\left(\frac{\beta_y}{1 + \beta_x} - \sin\theta
\right)x_s(t),
	\label{generation}
\end{equation}
where $a_0 = a\cos\theta$ is the incident wave amplitude in the boosted frame. 

The layer dynamics is determined by the equation
\begin{equation}
	\frac{d}{dt}x_s = \beta_x
	\label{motion}
\end{equation}
with the initial condition $x_s(t = 0) = 0$. By virtue of the ultrarelativistic
motion, the position of the layer may be found by assuming that the full
particle velocity is equal to the speed of light, i.e., $\beta^2_x + \beta^2_y =
1$. Equations (\ref{skining}) and (\ref{motion}) are then self-consistent and
the layer motion is described by a first-order nonautonomous ordinary
differential equation or by an autonomous system 
\begin{equation}
	\left\{
	\begin{array}{l}
	\displaystyle{
	\frac{du}{d\tau} = \frac{\left(u^2-1\right)\left(\sin\theta -u\right) \pm
4\cos^3\theta S^{-1}\left[1 - \eta^2\left(\sin\theta -
u\right)^2\right]^{1/2}}{\eta\left(u^2+1\right)} }\\
	\displaystyle{
	\frac{d\eta}{d\tau} = \frac{u^2-1}{u^2+1} }\\
	\end{array}
	\right.
	\label{system}
\end{equation}
for the variables $\eta(\tau) = x_s S/\left(2\cos^3\theta\right)$, $u(\tau) =
\beta_y/\left(1 - \beta_x\right)$, where $\tau = t S/\left( 2 cos^3\theta\right)$. 
The solution of Eq. (\ref{system}) depends on the two dimensionless variables
$S$ and $\theta$, and may be analyzed in the plane $\left\{\eta, u\right\}$, where we have two sheets
corresponding to the choice of sign in Eq.~(\ref{system}). 

\begin{figure}
\includegraphics[width=0.9\columnwidth]{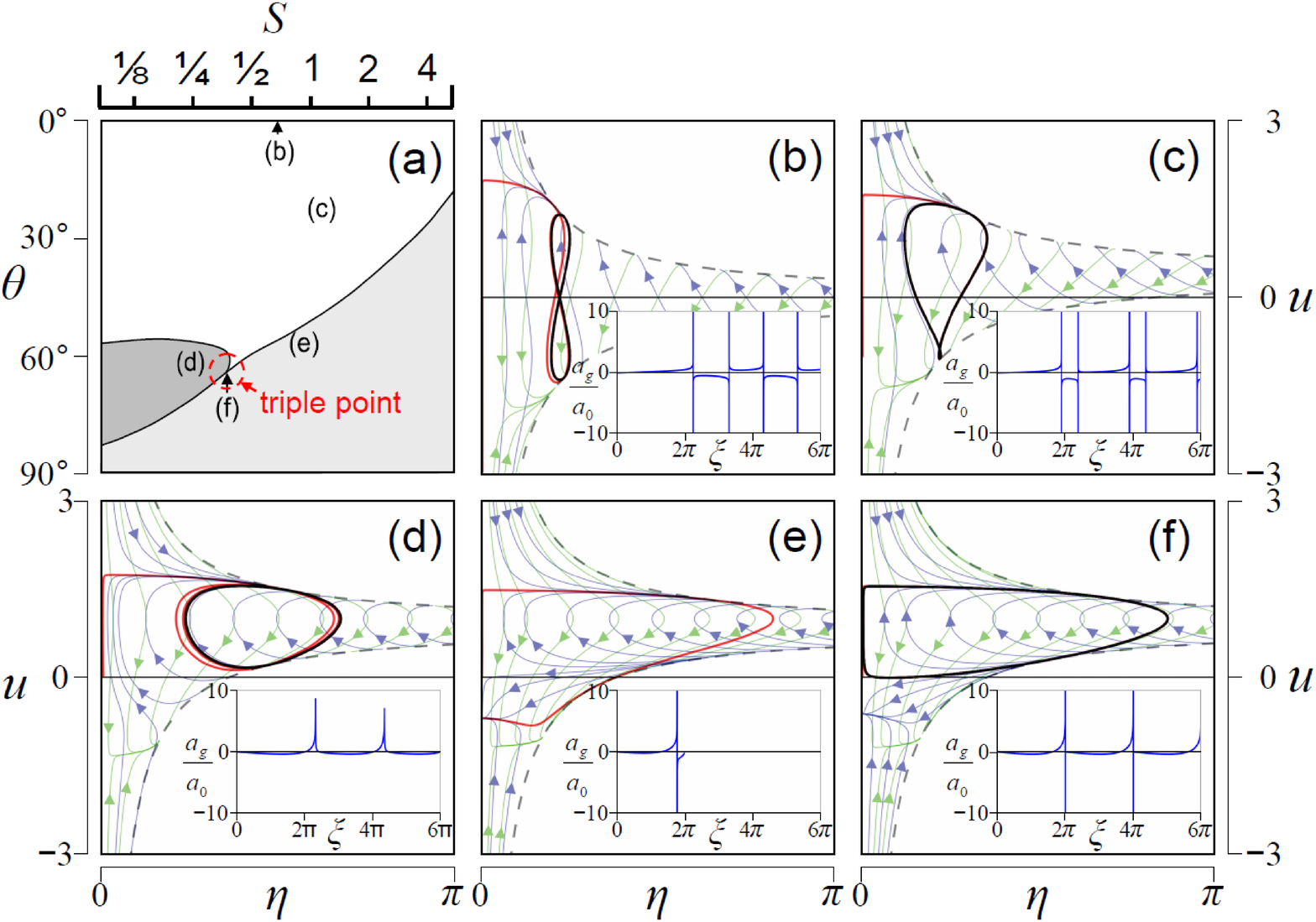}
\caption{(a) Regions on the plane of parameters $S, \theta$ corresponding to
qualitatively different forms of solution of Eq.~(\ref{system}); (b) $-$ (f)
form of solution on the  plane $\left\{\eta, u\right\}$ (red), its limit cycle (black), lines of phase
portrait for the $+$ sign (blue) and the $-$ sign (green) in Eq.
(\ref{system}), and the corresponding form of electromagnetic backradiation from the plasma.} \label{fig3}
\end{figure}

The topology of the phase plane is characterized by the existence of a stable
limit cycle (see Fig. \ref{fig3}). It is convenient to classify the form of the
solution by the number of zero-axis intersections of $u(\tau)$ or, equivalently, by
$\beta_y(t)$ in the $\eta > 0$ ($x_s > 0$) region. This takes place when $\beta_x \rightarrow
-1$ and Eq.\ (\ref{generation}) becomes singular, which corresponds to the
emission of the attosecond burst. There may occur either two such events (Fig.\
\ref{fig3}(c)), one (Fig.\ \ref{fig3}(e)), or none (Fig.\ \ref{fig3}(d)) in each
optical period. Accordingly, two bipolar, one bipolar, or one unipolar
attosecond pulse are generated. It is clear that we have the emission of two evenly spaced identical bursts
in the case of normal incidence (Fig.\ \ref{fig3}(b)). With increasing angle
$\theta$ the second burst either disappears due to amplitude decay down to zero
((c) $\rightarrow$ (e) transition) or two bipolar pulses merge into one unipolar
pulse as a result of convergence of their generation times ((c) $\rightarrow$
(d) bifurcation).
Our comprehensive numerical study indicates that the results obtained using the RES model are in a very good agreement with the PIC simulations for all values of $\theta$ and for $S<5$ down to RSIT (in particular, see Fig. \ref{fig2}(b), (c)). The RES model's applicability region is labeled "RES" in Fig. \ref{fig1}.

Note that by a simple modification of Eqs.\ (\ref{skining}) and
(\ref{generation}), the RES model can be easily generalized to take into account an arbitrary
plasma density profile, as well as arbitrary laser pulse shape and polarization.

In order to find the amplitude and duration of the pulse generated near $\beta_y
= 0$ one has to take into account the finite value of the relativistic factor
$\gamma$, which is the external parameter for the RES model and can be taken,
for example, from PIC simulation. Near the point $\beta_y = 0$,
$\beta_x\rightarrow -1$ the first term in the right-hand side of Eq.\
(\ref{generation}) dominates, so an analytical expression for the burst shape
and its spectrum can be written as follows
\begin{equation}
	a_g(\xi) = A_g f\left(\frac{\xi - \left.\xi\right|_{\beta_y =
0}}{\tau_g}\right),~
	I_k \propto \exp\left(-\frac{k}{\alpha \gamma^3}\right),
\label{gform}
\end{equation}
where $\alpha = \left.\frac{\partial \beta_y}{\partial t}\right|_{\beta_y = 0}$,
$f(\nu) = \frac{2\nu}{\nu^2 + 1}$, $A_g =  \frac{a_0 S \gamma
\left.x_s\right|_{\beta_y = 0}}{2 cos^3 \theta}$ is the pulse amplitude,
$\tau_g = \left(2\alpha \gamma^3\right)^{-1}$ is its characteristic duration,
and
$I_k$ is the intensity of the $k$th harmonic. The value of $\alpha$ is assessed
from the solution of the self-consistent system; hence, it depends only on
dimensionless parameters $S$ and $\theta$ and is independent of $\gamma$. This
means that pulse duration in the ultrarelativistic limit tends to zero as $\gamma^{-3}$. Note that in the RES model the spectrum decays exponentially with the characteristic scale $\alpha \gamma^3$ and, in contrast
to the OMM, has no region with a power-law decay. 
Thus, the "RES" region in Fig.~\ref{fig1} corresponds to a slower energy decay of the harmonics than the "OMM" region.

Depending on the thickness $l_s$ of radiating electron layer, the radiation can
be either coherent, for $l_s < \tau_g$, or incoherent, for $l_s > \tau_g$. In
the latter case, the assessment of the giant pulse amplitude requires taking
into account that only part of electrons radiates coherently. The radiating
electron layer thickness $l_s$ may be estimated assuming that, starting from the
time of maximum displacement, it decreases proportionally to the number of
electrons in the layer. Besides, as the electron velocity relative to the $\xi$
coordinate decreases from the speed of light to $c\left(1 + \beta_x\right)$, the
layer experiences an additional $2 \gamma^2$-fold compression. The layer
thickness $L_s$ at maximum displacement may be estimated from the equation of
balance between the force of light pressure and forces caused by charge
separation in the form $L_s \approx \sqrt[3]{2} \cos \theta S^{-1/3} a^{-2/3}$.
The resulting estimation of the adjustment factor is 
\begin{equation}
	C = \frac{\tau_g}{l_s} = \chi \frac{1}{\sqrt[3]{2}}\frac{1}{\alpha
\gamma}\frac{x_{max}}{\left.x_s\right|_{\beta_y = 0}}\frac{1}{\cos \theta}
S^{1/3} a^{2/3},
	\label{Correction}
\end{equation}
where $x_{max}$ is the maximum value of $x_s$, and $\chi$ is a dimensionless
constant of order unity that is needed to account for arbitrary choice of the
estimated values entering this expression. 

\begin{figure}
\includegraphics[width=0.9\columnwidth]{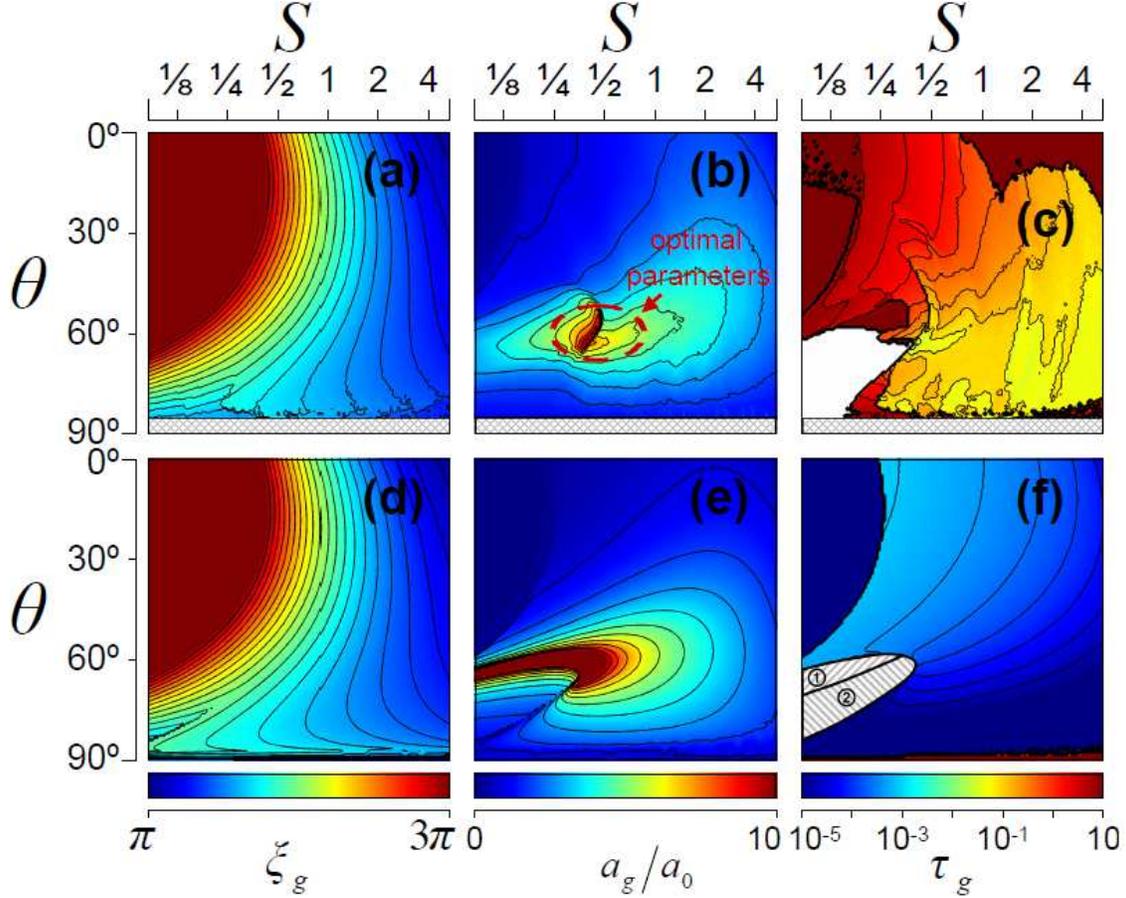}
\caption{Generation phase $\xi_g$, maximum amplitude $a_g/a_0$ and duration
$\tau_g$ of giant pulse obtained from the RES model with $\gamma = 30$ and $\chi
= 1/5$ (d, e, f) and from 1D PIC simulations (a, b, c) of semi-infinite plasma
obliquely irradiated by one optical cycle pulse with 1 $\mu$m wavelength and
amplitude $a = 191.1$, which corresponds to 10$^{23}$ W/cm$^2$ intensity. For
the PIC simulations, the pulse duration was assessed as a distance between
maximum and minimum electric field points, and the region of unipolar pulse
generation is given in white. On the (f) panel, the coherent generation and
monopolar pulse generation regions are marked (1) and (2), respectively.}
\label{fig4}
\end{figure}

The PIC simulations and the results of the numerical solution of Eqs.
(\ref{skining}), (\ref{motion}), including the adjustment factor
(\ref{Correction}), are compared in Fig. \ref{fig4}. We see that we have not
only qualitative, but also quantitative agreement for the burst amplitude and
the generation phase. 
A qualitative agreement can be seen for the burst duration and regions of unipolar
generation. The diagrams in Fig.~\ref{fig4}(b, c) allow us to distinguish the zone with the
center 
\begin{equation}
	\theta_g \approx 62^{\circ}, S_g \approx 1/2,
	\label{zone}
\end{equation}
and the boundaries $50^{\circ} < \theta < 70^{\circ}$, $1/4 < S < 1$ as the
region of the most powerful and short burst generation. This result may serve as
a guiding message for experimental implementation. The optimal parameters (\ref{zone}) correspond to the triple point in the $S$ and $\theta$ plane (Fig. \ref{fig3}(a, f)), which provides the longest time of irradiation and thus maximum amplitude of the generated attosecond pulse, as coherency plays the dominant role.

\begin{figure}
\includegraphics[width=0.9\columnwidth]{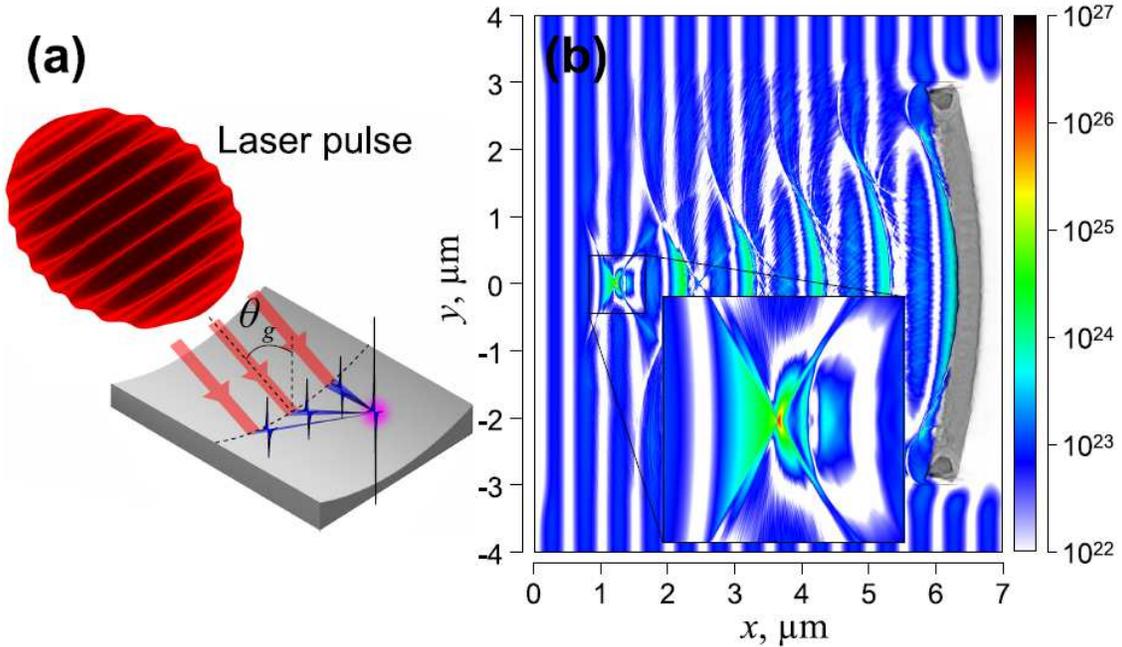}
\caption{(a) Schematic representation of the concept of groove-shaped target.
(b) Intensity distribution at focusing instant obtained from 2D PIC simulation:
a linearly polarized wave with intensity 10$^{23}$ W/cm$^2$ and wavelength 1
$\mu$m obliquely incident at an optimal angle $\theta_g = 62^{\circ}$ on a
parabolic groove-shaped target with density $1.065 \times$10$^{23}$ cm$^{-3}$,
which corresponds to $S = 1/2$.} \label{fig5}
\end{figure}

\paragraph{A focusing mechanism for attosecond pulses. ---} Based on the results obtained we propose a new concept of
extremely intense light generation at the level required for observation of the
QED effects. The idea is to focus the giant burst formed in the regime described
above by using a slightly grooved surface of the obliquely irradiated target at the optimal parameters (\ref{zone}), with the guiding line of the groove located in the plane of incidence (see Fig. \ref{fig5}(a)).
The PIC simulation of the proposed concept shows that the intensity
1.8$\times$10$^{26}$ W/cm$^2$ can be reached in the zone with the size of order
10 nm with a 10 PW laser pulse, as can be seen in Fig. \ref{fig5}(b). In the
laboratory frame the high field zone moves along the guiding line with the speed
$c/\sin\theta$. 
We note that the size of laser pulse along the transverse direction may be only a few wavelengths.
The data shown in Fig. \ref{fig4}
may be used to modify surface profile and target density to allow using a laser
pulse with a more complicated intensity profile in the transverse direction.

\paragraph{Conclusion. ---} In this work we studied the giant pulse generation
process at oblique irradiation in an overdense plasma by a relativistically
strong laser pulse. The model of \textit{relativistic electronic spring} was
developed, providing a qualitative and, for some characteristics, also a fairly
good quantitative description. The parameters of the most powerful burst
generation (\ref{zone}) were determined. A new concept of a groove-shaped
target for high electromagnetic field generation aimed at obtaining the QED
effects by means of upcoming laser sources was proposed and confirmed by PIC
simulation.

\paragraph{Methods. ---} For the numerical studies we used a particle-in-cell approach. The data shown in Fig. (\ref{fig1}), (\ref{fig2}) and (\ref{fig4}) were obtained using 1D simulations in the moving frame (to take into account the oblique incidence of the laser pulse), implying a plasma flux in the transverse direction. For Fig. (\ref{fig5}) we have carried out a 2D simulation in the frame moving along the guiding line of the groove. 
Performing the PIC simulation in the moving frame implies that
the transverse size of the laser pulse is fairly large compared to the
wavelength: However, this is a rather weak restriction. A fully relativistic
parallel FFT based PIC code ("ELMIS" \cite{ELMIS}) have been used; an 8 $\mu$m $\times$ 8 $\mu$m
region is represented by 8192$\times$8192 cells. The plasma ions are taken to be
Au$^{6+}$, and each target cell contains approximately 100 virtual particles of each
species. The time step is 1.07$\times$10$^{-2}$ fs, and the laser pulse front has a
sine-squared profile with a two wave periods duration.

\paragraph{Acknowledgements. ---} This research was supported by the Presidium of RAS, the RFBR (Grant No. 09-02-12322-ofi\_m), the Presidential Council on Grants of the Russian Federation (Grant No. 3800.2010.2), the European Research Council (Grant No.\ 204059-QPQV), and the Swedish Research Council (Grant No.\ 2007-4422). We acknowledge the Joint Supercomputer Center of RAS and the Swedish National Infrastructure for Computing (SNIC) for the provided supercomputer sources.

\end{document}